\documentclass[proof]{WileyASNA-v1}

\articletype{Article Type}%

\accepted{18 October 2021}

\begin{document}

\title{Jet-triggered star formation in young radio galaxies}

\author[1]{Chetna Duggal*}
\author[1]{Christopher O'Dea}
\author[1]{Stefi Baum}
\author[2,3]{Alvaro Labiano}
\author[4,5]{Raffaella Morganti}
\author[6]{Clive Tadhunter}
\author[7]{Diana Worrall}
\author[8]{Grant Tremblay}
\author[9]{Daniel Dicken}
\author[10]{Alessandro Capetti}

\authormark{DUGGAL \textsc{et al}}

\address[1]{\orgdiv{Department of Physics and Astronomy}, \orgname{University of Manitoba}, \orgaddress{\state{Manitoba}, \country{Canada}}}

\address[2]{\orgdiv{Centro de Astrobiolog\'ia (CAB, CSIC-INTA)}, \orgname{ESAC Campus, E-28692 Villanueva de la Cañada}, \orgaddress{\state{Madrid}, \country{Spain}}}

\address[3]{\orgdiv{Telespazio UK for the European Space Agency (ESA)}, \orgaddress{\state{ESAC}, \country{Spain}}}

\address[4]{\orgdiv{Kapteyn Astronomical Institute}, \orgname{University of Groningen}, \orgaddress{\state{9700 AB Groningen}, \country{The Netherlands}}}

\address[5]{\orgdiv{ASTRON, the Netherlands Institute for Radio Astronomy}, \orgaddress{\state{Postbus 2, NL-7990 AA Dwingeloo}, \country{The Netherlands}}}

\address[6]{\orgdiv{Department of Physics $\&$ Astronomy}, \orgname{University of Sheffield}, \orgaddress{\state{Sheffield}, \country{UK}}}

\address[7]{\orgdiv{H.H. Wills Physics Laboratory}, \orgname{University of Bristol}, \orgaddress{\state{Bristol}, \country{UK}}}

\address[8]{\orgdiv{Harvard-Smithsonian Center for Astrophysics}, \orgaddress{\state{Cambridge, MA}, \country{USA}}}

\address[9]{\orgdiv{AIM, CEA, CNRS}, \orgname{Universit\'e Paris-Saclay, Universit\'e Paris Diderot, Sorbonne Paris Cit\'e}, \orgaddress{\state{F-91191 Gif-sur-Yvette}, \country{France}}}

\address[10]{\orgdiv{INAF - Osservatorio Astrofisico di Torino}, \orgaddress{\state{Pino Torinese}, \country{Italy}}}

\corres{*Chetna Duggal, University of Manitoba
Winnipeg, MB R3T 2N2, Canada. \email{duggalc@myumanitoba.ca}}


\abstract{Emission in the ultraviolet continuum is a salient signature of the hot, massive and consequently short-lived, stellar population that traces recent or ongoing star formation. With the aim of mapping star forming regions and morphologically separating the generic star formation from that associated with the galaxy-scale jet activity, we obtained high-resolution UV imaging from the \textit{Hubble Space Telescope} for a sample of nine compact radio sources. Out of these, seven are known Compact Steep Spectrum (CSS) galaxies that host young, kiloparsec-scale radio sources and hence are the best candidates for studying radio-mode feedback on galaxy scales, while the other two form a control sample of larger sources. Extended UV emission regions are observed in six of the seven CSS sources showing close spatial alignment with the radio-jet orientation. If other mechanisms possibly contributing to the observed UV emission are ruled out, this could be evidence in support of jet-triggered star formation in the CSS phase of radio galaxy evolution and in turn of the `positive feedback' paradigm of host-AGN interaction.}


\jnlcitation{\cname{%
\author{Duggal, C.}, 
\author{O'Dea, C.}, 
\author{Baum, S.}}, \cyear{2021}, \cjournal{Astron. Nachr.} \cvol{1}. https://doi. org/10.1002/asna.20210054}


\maketitle



\section{introduction}\label{sec1}


\begin{center}
\begin{table*}[t]%
\caption{List of observed targets\label{tab1}}
\centering
\begin{tabular*}{500pt}{@{\extracolsep\fill} c c c c c c c c c }
\toprule
\textbf{Source} & \textbf{Catalogue} & \textbf{\textit{z}} & \textbf{Radio size} & \textbf{Proj. linear} &\textbf{Sample}\tnote{$^\dagger$} & \textbf{FUV} & \textbf{V}  & \textbf{UV detection}$^{\ast}$ \\ 
 & \textbf{name} & & (arcsec) & \textbf{size} (kpc) &  & \textbf{filter} &  \textbf{filter} &  \\
\midrule
B0258+35 & NGC 1167 & 0.017 & 3.8  & 1.32  &  G05 & F225W & F621M & \\
B1014+392$^\otimes$ & 4C 39.29 & 0.536 & 6.1  & 39.03  & F01 & F336W & F763M & \\
B1025+390 & 4C 39.32  & 0.361 & 3.2 & 16.28 & F01 & F336W & F763M & \checkmark \\
B1037+30 & 4C 30.19 & 0.091 & 3.3  & 5.63 &  G05 & F225W & F621M & \checkmark \\
B1128+455 &  & 0.404 & 0.9  &  4.91 & F01 & F336W & F763M & \checkmark \\
B1201+394 &  & 0.445 & 2.1 & 12.14  & F01 & F336W & F845M & \checkmark \\
B1203+645 & 3C268.3  & 0.371 & 1.4  & 7.25 &  O98 & F336W & F763M & \checkmark \\
B1221-423 & PKS 1221-42  & 0.171 & 1.5  & 4.40  & B06 & F275W & F689M & \checkmark \\
B1445+410$^\otimes$ &  & 0.195 & 8.1 & 26.41 & F01 & F275W & F689M & \\
\bottomrule
\end{tabular*}
\begin{tablenotes}
\item[$^\otimes$] The non-CSS control sample (projected linear size of radio source > 20 kpc). $^\dagger$Samples: G05 \citep{2005A&A...441...89G} = low power CSS; F01 \citep{2001A&A...369..380F} = moderate power CSS; O98 (O'Dea 1998 = \cite{1997A&A...325..943S} plus \cite{1990A&A...231..333F}); B06 \citep{2006AJ....131..100B} = southern 3C equivalent. $^\ast$The subset of target sources with a definite detection in the UV band is signified with \checkmark marks.
\end{tablenotes}
\end{table*}
\end{center}

\hspace{-3mm}Compact Steep-Spectrum (CSS) sources are a subclass of compact radio-luminous active galactic nuclei (AGN) characterized by projected linear sizes from $\sim500$ parsecs (pc) to 20 kiloparsecs (kpc), and steep radio spectra ($\alpha \geq 0.5$, where flux density, S $\propto \nu^{- \alpha}$) that tend to peak at radio frequencies below about $\sim 400$ MHz \citep{2021A&ARv..29....3O}. Gigahertz-peaked spectrum sources and High-Frequency Peakers, collectively referred to as Peaked Spectrum (PS) sources, along with the Compact Symmetric Objects (CSOs) encompass the other types of compact radio sources that typically show projected linear sizes smaller than 500 pc. 

While the small-scale extent of jet emission in compact sources could be attributed to episodic or transient activity (e.g., \cite{2010MNRAS.408.2261K}), confinement to host galaxy atmospheres (e.g., \cite{1984Natur.308..619W}), or radio-enhancement in intrinsically weaker sources (e.g., \cite{1991ApJ...373..325G, 1998PASP..110..493O, 2011MNRAS.412..960T, 2012ApJ...745..172D}) due to interactions with dense interstellar medium (ISM) of host galaxies, it has been argued that a section of this compact radio source population is likely to be intrinsically small, young radio sources \citep{2021A&ARv..29....3O} that would eventually grow and develop into large-scale radio galaxies and quasars (e.g. FR I/II galaxies, \cite{1974MNRAS.167P..31F}) and thus represent an early stage of radio galaxy evolution \citep{1995A&A...302..317F, 1996ApJ...460..634R, 1998PASP..110..493O}.

This period of infancy is where the radio source is likely to interact vigorously with the surrounding gas, driving shocks through the ISM, and triggering or boosting star formation in the host galaxy. Given the lifetimes of hot, massive stars, this so-called positive feedback would happen entirely during the CSS phase when the nuclear jets are confined to galaxy extents. Hence, jet-induced star formation is expected to be a key signature of radio-mode feedback on galaxy scales \citep{1989MNRAS.239P...1R, 2014ApJ...796..113D, 2012MNRAS.425..438G, 2017ApJ...850..171F}.

The phenomenon of jet-driven star formation has been supported by observational evidence in some extended radio sources (e.g. \cite{2006ApJ...647.1040C, 2015A&A...574A..34S, 2020A&A...639L..13N, 2020MNRAS.499.4940Z}). The observation of similar effects in the context of young radio sources, however, has been limited. Since PS/CSO sources have sizes too small to be sufficiently resolved on the scale of the radio source in the optical, IR $\&$ UV bands with current facilities, CSS sources arguably provide the best opportunity to study the effects of galactic-scale jet activity on host ISM. 

Extended emission-line regions in CSS galaxies have been often observed to be co-spatial and aligned with the radio source \citep{2000AJ....120.2284A, 2008ApJS..175..423P}. Such an alignment effect is generally attributed to outflows driven by the kpc-scale radio source (see \cite{2021A&ARv..29....3O} and the references therein). Therefore, the observation of extended UV emission with a similar close spatial alignment to radio source orientation in these compact sources is a justifiable argument in favour of the radio-jet feedback paradigm. 
\newline

\hspace{-3mm}Following our pilot study that detected extended UV light aligned with the radio source in 2 out of 3 CSS sources (3C 303.1 and 1814-637;  \cite{2008A&A...477..491L}), we went on to conduct a UV imaging survey using the \textit{Hubble Space Telescope} (HST) of a larger sample of nine sources. We also obtained short optical images which allow us to confirm that the UV light is indeed due to newly-formed stars as well as to determine the spatial distribution of the older stellar population.
\vspace{8mm}

\begin{figure}[t]
\centerline{\includegraphics[scale=0.3]{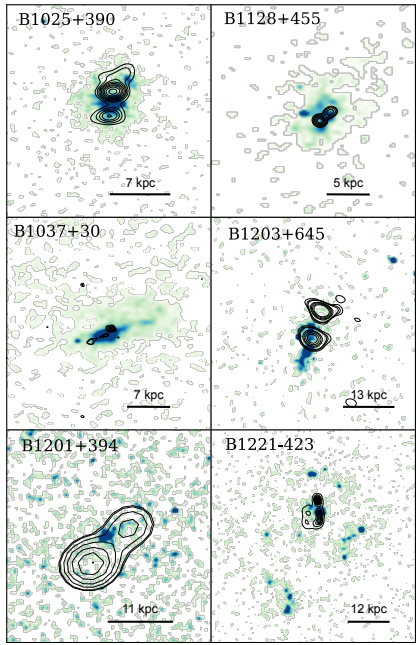}}
\caption{HST UV continuum images overlaid with contours from VLA radio data. 
\label{fig1}}
\end{figure} 

\section{The sample}\label{sec2}

The nine compact radio galaxies selected for our study (listed in Table 1) were chosen 
such that they are at relatively nearby redshifts ($z \lesssim 0.6$), thus eliminating strong effects due to evolution with cosmic time. Their radio source sizes range from 1$-$8" so as to have good resolution with the HST/WFC3 imager along the radio source. These target sources were drawn from well-defined samples compiled in previous studies (Column 6, Table 1), seven of which are known CSS radio sources while the other two represent a control sample of larger sources. 

\section{observations}\label{sec3}

We carried out high-resolution ($\sim 0.05''$/pix) imaging of the nine compact radio sources with the UVIS channel on the HST's \textit{Wide Field Camera 3} (WFC3) in the optical ($6000-8500$ \AA) and ultraviolet ($2000-3500$ \AA) bands. 
The filter selection (Table 1) was based on our requirement of imaging the UV and optical continua free of bright emission lines. In addition, archival radio maps of the target CSS sources from the \textit{Very Large Array} (VLA) were used to determine the spatial relationship between the extended UV emission regions and the radio source.

The HST/WFC3 imaging data were reduced using a combination of basic calibrations performed as part of the standard HST \textit{calwf3} pipeline -- i.e., bias $\&$ dark current subtraction, flat-fielding, linearity and charge transfer efficiency corrections -- with manual post-pipeline reprocessing using the Drizzlepac software package~\citep{2021drzp} for improved cosmic-ray rejection, geometric distortion correction and customization of image alignment and dithering to produce the final drizzled images.

\section{uv morphology $\&$ jet-driven star formation}\label{sec4}

\begin{figure}[t]
\centerline{\includegraphics[scale=0.3]{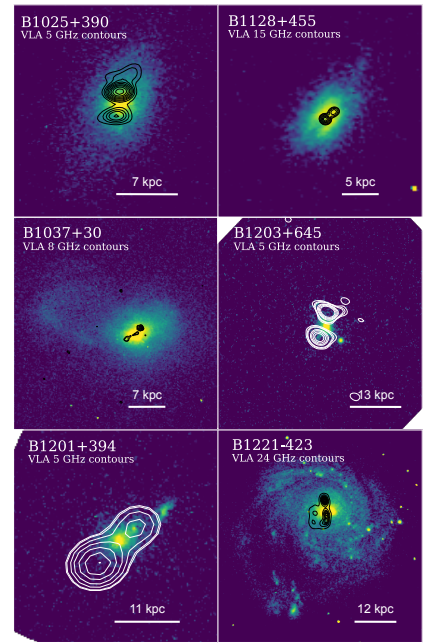}}
\caption{HST/WFC3 optical continuum images overlaid with VLA radio observations for the six UV-detected CSS targets. 
\label{fig2}}
\end{figure} 

Our aim was to study the spatial relationship between star formation and powerful radio-jet emission by morphologically separating the generic star formation due to gas infall, from that associated with the kpc-scale radio sources and hence probably triggered due to jet activity. 

\hspace{-3mm}With our near-UV band imaging, we detected extended UV continuum emission in 6 out of the 9 compact radio sources in our sample (Figure 1). While the bright UV emission clustered around the core regions is more likely to be due to the central AGN, the extended and clumpy structures in the UV are more likely to be star-forming regions. A visual comparison of the UV and optical band images of the CSS host galaxies (Figures 1 $\&$ 2) shows the distribution of the younger stellar populations as evident from UV continuum emission, relative to the general galaxy morphology.  

The extended UV light exhibits strong alignment with the radio-jet axes in atleast 5 out of our target sources (a higher resolution radio image is needed to confirm the presence of this effect in B1201+394, see Figure 1), which suggests a dynamic feedback relationship between jet activity and the host galaxy ISM. This might be strong evidence in support of jet-induced star formation, if the clumpy/extended UV continuum emission observed in these CSS sources could indeed be attributed to star-forming regions. 

\section{discussion}\label{sec5}

While our data show strong evidence in support of jet-driven star formation during the Compact Steep Spectrum phase of radio evolution, further investigation is required to confirm the origin of the observed UV emission. While the extended UV- continuum emission co-spatial with the radio source is likely to have originated in star formation due to dense shocked gas in the galactic ISM, the UV light extending beyond the radio source could be a consequence of past merger events in some of the sources, as evident from their perturbed optical morphologies. Another possibility is a repetitive or restarted activity scenario where the UV emission on larger scales was caused by an older radio source, larger in extent than the present, nascent radio source. 
Additionally, there might be contamination from AGN-related components, i.e., scattered UV light from the central AGN and/or nebular continuum emission from AGN-ionized emission regions (e.g., \cite{1995MNRAS.273L..29D,1997ApJ...476..677C, 2002MNRAS.333..211W, 2002MNRAS.330..977T}). These, if present, would be apparent in polarimetric UV imaging and spectroscopic observations of the CSS galaxies, respectively. We plan to address these questions as part of our ongoing and future work. 

\section*{Acknowledgments}

The research of C. Duggal, C. O'Dea, and S. Baum is supported by the Natural Sciences and Engineering Research Council (NSERC) of Canada. This research made use of Astropy,\footnote{http://www.astropy.org} a community-developed core Python package for Astronomy. 

\bibliography{Wiley-ASNA}%


\vspace{-3mm}
\section*{Author Biography}

\begin{biography}{}
{\textbf{Chetna Duggal} is a PhD student at the University of Manitoba.
Her doctoral thesis is focused on studying the signatures of host-AGN interaction in powerful radio galaxies using a multi-wavelength approach.}
\end{biography}

\end{document}